# Spin texture induced by non-magnetic doping and spin dynamics in 2D triangular lattice antiferromagnet *h*-Y(Mn,Al)O$_3$

Pyeongjae Park[1,2,3,†], Kisoo Park[2,3,†], Joosung Oh[2,3], Ki Hoon Lee[2,3,4,5], Jonathan C. Leiner[2,3,6], Hasung Sim[2,3], Taehun Kim[1,2,3], Jaehong Jeong[1,2,3], Kirrily C. Rule[7], Kazuya Kamazawa[8], Kazuki Iida[8], T. G. Perring[9], Hyungje Woo[9,10], S.-W. Cheong[11], M. E. Zhitomirsky[12], A. L. Chernyshev[13] & Je-Geun Park[1,2,3,*]

[1]*Center for Quantum Materials, Seoul National University, Seoul 08826, Republic of Korea*

[2]*Center for Correlated Electron Systems, Institute for Basic Science, Seoul National University, Seoul 08826, Republic of Korea*

[3]*Department of Physics and Astronomy & Institute of Applied Physics, Seoul National University, Seoul 08826, Republic of Korea*

[4]*Center for Theoretical Physics of Complex Systems, Institute for Basic Science, Daejeon 34126, Korea*

[5]*Department of Physics, Incheon National University, Incheon 22012, Korea*

[6]*Physik-Department, Technische Universität München, D-85748 Garching, Germany*

[7]*Australian Nuclear Science and Technology Organisation, Lucas Heights, 2234 New South Wales, Australia*

[8]*Comprehensive Research Organization for Science and Society (CROSS), Tokai, Ibaraki 319-1106, Japan*

[9]*ISIS Pulsed Neutron and Muon Source, STFC Rutherford Appleton Laboratory, Didcot, Oxfordshire, OX11 0QX, United Kingdom*

[10]*Department of Physics, Brookhaven National Laboratory, Upton, New York 11973, USA*

[11]*Department of Physics and Astronomy, and Rutgers Center for Emergent Materials, Rutgers University, Piscataway, New Jersey 08854, USA*

[12]*Université Grenoble Alpes, CEA, IRIG, PHELIQS, 38000 Grenoble, France*

[13]*Department of Physics and Astronomy, University of California, Irvine, California 92697, USA*

* Corresponding author: jgpark10@snu.ac.kr

**Novel effects induced by nonmagnetic impurities in frustrated magnets and quantum spin liquid represent a highly nontrivial and interesting problem[1-4]. A theoretical proposal[5-7] of extended modulated spin structures induced by doping of such magnets, distinct from the well-known skyrmions has attracted significant interest. Here, we demonstrate that nonmagnetic impurities can produce such extended spin structures in *h*-YMnO$_3$, a triangular antiferromagnet with noncollinear magnetic order. Using inelastic neutron scattering (INS), we measured the full dynamical structure factor in Al-doped *h*-YMnO$_3$ and confirmed the presence of magnon damping with a clear momentum dependence. Our theoretical calculations can reproduce the key features of the INS data, supporting the formation of the proposed spin textures. As such, our study provides the first experimental confirmation of the impurity-induced spin textures. It offers new insights and understanding of the impurity effects in a broad class of noncollinear magnetic systems.**

In many geometrically frustrated magnets, a weak perturbation can potentially induce new competing ground states, leading to an extremely rich phase diagram. A particularly important but less explored case is that of the frustrated system with defects, i.e., nonmagnetic impurities, which are ubiquitous in real-world materials. Combined with the geometrical frustration, impurities can introduce several nontrivial effects in a wide variety of frustrated magnets. For example, extensive search for a genuine quantum spin liquid (QSL) state in frustrated magnets has been consistently plagued by the impurity issues, e.g., herbertsmithite and YbMgGaO$_4$ as the S=1/2 kagome[3,8,9] and triangular[4,10,11] QSL candidates, respectively. It has been theoretically pointed out that impurities themselves can induce continuum-like excitations in geometrically frustrated magnets, which is difficult to distinguish from the genuine signature of a QSL state. Other examples are a rich phase diagram due to impurities in a triangular lattice antiferromagnet[12] (TLAF), an impurity-induced spin-glass state in the frustrated spinel[13-15] and spin ice[16,17] systems, suppression of the QSL state due to impurities[2], and a disorder-induced classical[18,19] and quantum[1] QSL state, to name a few.

One of the salient features resulting from a nontrivial interplay between the geometrical frustration and disorder is a significant modification of spins' directions near impurities. In frustrated magnets, a vacancy gives rise to a partial relief of local frustration around it, which leads to a sizable reorientation of spins proximate to the impurity site (Fig. 1a). In a coplanar magnetic ground state with a well-defined easy plane, the reorientation only consists of in-plane components. The reorientation at the six nearest sites points toward the opposite direction to the effective impurity moment, leading to a screening of the effective impurity moment. Subsequently, spins at the next-nearest sites are affected by the nearest sites' reorientation, leading again to the canting toward the direction opposite to the former reorientation (See Fig. S1 for demonstration of the spatial structure of the spin reorientation). Such a spatially-correlated pattern of the spin reorientation acts as a partial screening of the effective impurity moments, resulting in an uncompensated fractional impurity moment[5], among other phenomena.

Moreover, the spins' canting angle decays algebraically to the distance from the vacancy[5,20,21] (Fig. 1b), resulting in a formation of an extended spin object around the vacancy called *spin texture*[5-7]. This impurity-induced texture can affect the spin dynamics and the ground state of the original system, which determines the thermal and spin transport properties of the material. Therefore, it may have a broad applicability to several fields in magnetism, not to mention its importance as a new way of generating large spin objects analogous to a skyrmion. However, despite such outstanding interests, there has been no experimental demonstration of the impurity-induced spin textures as of yet.

*h*-YMn$_{1-x}$Al$_x$O$_3$ is a unique model system to study dilution effects in a frustrated magnet, where non-magnetic Al$^{3+}$ ions are doped into the Mn$^{3+}$ triangular lattice[22,23]. Pure *h*-YMnO$_3$ has noncollinear 120° magnetic order below $T_N$ = 74 K due to the geometrical frustration. The spin dynamics of *h*-YMnO$_3$ have been well established by the previous studies[24-31], which used a simple model Hamiltonian (Eq. 1) with some additional terms from magnon-magnon/phonon coupling. Most importantly, increasing Al concentration x up to 0.2 does not change the crystalline symmetry but only reduces the antiferromagnetic transition temperature (Fig. 1c) and the magnetic moments of Mn$^{3+}$ gradually, enabling a systematic study of the dilution effect on the magnetism of a triangular lattice antiferromagnet (TLAF)[23]. While a large single crystal of *h*-YMn$_{1-x}$Al$_x$O$_3$ suited for inelastic neutron scattering (INS) is available[23], previous studies on this material focused only on the crystal structure and its bulk properties[23,32,33]. Here, we report the full spin dynamics of *h*-YMn$_{1-x}$Al$_x$O$_3$ studied by INS and model calculations, which provide the first experimental evidence of the impurity-induced spin texture.

Fig. 2a-c show the energy-momentum ($Q$) slices from the INS data of 0%, 10%, and 15% Al-doped $h$-YMnO$_3$ single crystals along the high-symmetric lines (see Fig. 1d), which demonstrate the influence of the Al doping. In contrast to the clear INS spectra of the pristine $h$-YMnO$_3$, the INS spectra of the Al-doped samples show broad magnon signals as expected for magnetic systems with impurities. Interestingly, however, the observed energy linewidth broadening is not uniform over the Brillouin zone but $Q$-dependent. A stack of constant $Q$-cuts along the B−C ([K −K 0] direction) in Fig. 3a-c supports this statement: near the magnetic zone center (C point), the linewidth of the magnon peaks is hardly changed with doping, whereas the magnon peaks near the zone boundary (B point) undergo a drastic broadening. For a more transparent demonstration of the $Q$-dependent energy linewidth broadening, see Fig. S5. For a deeper understanding of the $Q$-dependent linewidth broadening, we analyzed the half-width at half-maximum (HWHM) of the magnon peaks (see Methods) over the full Brillouin zone (Fig. 3d-f), which was averaged for all magnon modes at each $Q$-point. We note that there is no clear $Q$-dependence of the HWHM in the pure $h$-YMnO$_3$ (Fig. 3d). In $h$-Y(Mn,Al)O$_3$, however, the HWHM steadily increases when the magnon branches get closer to the zone boundary, and reaches its maximum near the B point (equivalent to the M point in the reciprocal space of a triangular lattice). Another noticeable feature is the $Q$-dependent magnon energy renormalization, manifested in the increasingly downward shift of the magnon dispersion along the A-B direction (Fig. 2a-c). Such $Q$-dependent energy linewidth broadening and renormalization of magnons may imply the presence of something beyond a simple dilution effect, as a point-like scattering potential made by an impurity is known to induce a $Q$-independent behavior of magnons generally[7,34,35].

To explain the observed features, we have modeled the spin Hamiltonian of a diluted triangular lattice with randomly distributed vacancies (see Methods for details). We assumed that Al doping does not change the parameters of the spin Hamiltonian, which is plausible considering the simple linear relation between x and $T_N$ (Fig. 1c)[36]. The spin dynamics of pure $h$-YMnO$_3$ can be described by the following spin Hamiltonian suggested by the previous study[24]:

$$H_{spin} = J_1 \sum_{\langle ij \rangle} \mathbf{S}_i \cdot \mathbf{S}_j + D_1 \sum_i (s_i^z)^2 + D_2 \sum_i (\mathbf{S}_i \cdot \hat{n})^2, \tag{1}$$

where $J_1$ denotes the coupling constant of nearest-neighbor super-exchange interaction, and $D_1$ and $D_2$ are the size of the easy-plane anisotropy and the local easy-axis anisotropy along the direction of each spin ($\hat{n}$) in the 120° magnetic structure, respectively. Note that we assumed an ideal triangular lattice without trimerization present in $h$-YMnO$_3$ to focus on the impurity effects for this work, as the change due to the trimerization would be marginal from the viewpoint of its spin dynamics. For model calculations with the trimerization effect, see the Supplementary Information (Fig. S2). We adopted the parameters from ref. [24]: $J_1$ = 2.5 meV, $D_1$ = 0.28 meV, and $D_2$ = −0.02 meV. Note that both $D_1$ and $D_2$ contribute to forming the two gaps at 5 and 2 meV, respectively, leading to the double peak structure at the C point in both pure and Al-doped YMnO$_3$ (Fig. 2).

First, we tested the ground state modification by a single impurity (see Methods) and compared it with our previous theoretical work results. Fig. 1b shows the canting angle of spins $|\delta\Theta(r)|$ as a function of the distance from the impurity site ($r$) on a logarithmic scale, both with and without the easy-axis anisotropy $D_2$. For the directions of the spin canting, see Fig. S1. Without the easy-axis anisotropy ($D_2$=0), the canting angle follows asymptotically the algebraic decay law, $1/r^3$, indicated by the guide line[5]. The long-range spin texture is formed with the canting angle depending on the distance $r$ as well as on the sublattice number[12]. At large distances ($r$>30), the numerical data are affected by finite-size effects. Turning on $D_2$, the canting angle decreases faster with distance, as expected for anisotropic models, but the spin texture remains almost intact at intermediate distances.

Using the relaxed spin configuration, including the spin texture and Eq. 1, we calculated the INS

cross-section of $YMn_{1-x}Al_xO_3$ at the level of linear spin-wave theory (LSWT); see Methods for the details. Fig. 2d-f and 3a-c show the results of the calculations convoluted with the instrumental resolution for x=0, 0.1, and 0.15, which are in excellent agreement with our INS data (for the results without the resolution convolution, see Fig. S7 and S8). These results suggest that despite the simplicity of LSWT, our model calculation has successfully captured the unusual features observed in the data: the $Q$-dependent energy linewidth broadening and the increase of the downward shift of the magnon dispersion along the A-B direction. Note that the discrepancy between the magnon dispersion along A-B in Fig. 2a and 2d (x=0) is due to the effect of magnon-phonon coupling present in $h$-$YMnO_3$ (ref. [24]), which was not considered in our model calculation.

Such a $Q$-dependent magnon lifetime in a diluted noncollinear magnet has already been suggested by a previous theoretical study[7], which argued that the spin texture gives rise to $Q$-dependent magnon scattering. According to ref. [7], the impurity-induced spin texture generates a spatially-dispersed effective magnetic field around an impurity (unlike a point-like potential made by the impurity itself), which acts as a potential for the magnon Umklapp scattering. Importantly, this scattering makes the scattering rate strongly $Q$-dependent as exactly found in our experiments. As the effect mentioned above is already embedded in our model calculations, the $Q$-dependent behavior observed in both the data and the calculations would be from the magnons' scattering on the spin textures. Note that similar $Q$-dependent magnon scattering was also suggested in the system with skyrmions, where the effective magnetic field from its topological texture acts as a scattering potential[37]. To verify that the observed behavior is unique characteristics of a frustrated magnet, we contrast it with a diluted non-frustrated square-lattice antiferromagnet's spin dynamics. As shown in Fig. S9, impurities in a square lattice create a flat localized mode, consistent with the previous INS study on a perovskite fluoride $K(Co,Mn)F_3$ (ref. [38]). However, one does not see any noticeable $Q$-dependent energy linewidth broadening or renormalization when increasing doping. Such distinctive difference comes from the absence of geometrical frustration (and, therefore, the absence of spin textures) in the square lattice antiferromagnet as opposed to the TLAF. Further theoretical analysis is needed for a comprehensive understanding of the specifics of the observed $Q$-dependence, such as why the most drastic changes in the spectrum due to dilution occur in the vicinity of the B point, which we leave for further studies.

To explicitly examine the spin textures' role in the spin dynamics of $h$-Y(Mn,Al)O$_3$, we have also calculated the INS cross-sections of a diluted triangular lattice without the feedback of the vacancy, i.e., without the spin texture. To perform such calculation, we artificially forced the spins to retain the 120° magnetic order for the calculation, which is similar to the approach used in ref. [7]. It amounts to neglecting the feedback of the impurity onto the host spins, so that no texture is created. This allows us to separate scattering effects of the conventional dilution from the effects associated with extended textures. Fig. 4a-b show the calculated magnon spectra of 10% Al-doped $h$-YMnO$_3$ with and without the spin texture. While there exist a couple of slight differences when comparing the results in Fig. 4a and 4b over the full Brillouin zone, we found a particularly large difference in the intensity of the 5 meV mode near the C point. Complementary calculations without the spin texture confirmed that the 5 meV mode becomes strongly suppressed due to the non-magnetic impurities (Fig. S10). In comparison, there is no noticeable suppression of the 5 meV mode in our experimental data and the theoretical calculation with the spin texture (Fig. 4c). This result implies that the in-plane spin texture formation is a key factor in retaining the 5 meV mode's stability against the non-magnetic impurity.

To further understand the origin of the significant difference at the C point near 5 meV between Fig. 4a and 4b, we analyzed the eigenvector of the 5 meV magnon mode at the C point in pure $h$-YMnO$_3$ (Fig. S10). This exercise allows us to examine why it becomes particularly susceptible to the vacancy at the C point as a perturbation. As expected, an out-of-plane motion is dominant for the spin precession of the 5 meV mode, in accordance with the fact that this mode is gapped by the easy-plane anisotropy $D_1$. However, we also found some finite in-plane precession components in its eigenvector. Notably, the in-plane precession of the six spins nearest to a specific site for the 5 meV magnon mode is almost identical to the effect of the spin canting of six spins due to the vacancy (see Fig. S10). In other words, the 5 meV eigenmode will undergo significant energy (eigenvalue) variation due to the vacancy, indicating its sensitivity to the vacancy as a perturbation. Moreover, since the spin precession is synchronized over the triangular lattice due to a zero magnon wave-vector at the C point, the effects of this sensitivity will be amplified, leading to an ill-defined spectrum of the 5 meV mode seen in Fig. 4b. While further confirmation of whether such a minor portion of the spin precession can result in the significant change is required, these results imply that spin textures may play an important role in explaining the diluted noncollinear magnets' magnetic excitations.

Although our INS data of $YMn_{1-x}Al_xO_3$ together with the model calculations have provided some valuable insight about the spin texture, we would like to note that such an approach would be somewhat close to the indirect examination. Therefore, further measurements to directly detect the spin texture in real space will be of great help to deeper understanding of it. For instance, small angle neutron scattering (SANS) may give further hidden information about its spatial correlation, which is the key feature of the spin texture.

In summary, our work presents a unique experimental study on the energy and momentum-resolved spin dynamics of a diluted frustrated magnet. It contributes to answering the critical fundamental problem of the nontrivial impurity effects in frustrated magnets. Furthermore, our results provide the first experimental confirmation of the impurity-induced spin textures, which have been long-advocated theoretically. We demonstrate that generating a spin texture can be easily achieved using frustrated noncollinear magnets. It may also be conceivable that this giant spin texture can be manipulated and so used as potential applications as done for skyrmion.

## Methods

**Sample preparation.** $h$-YMn$_{1-x}$Al$_x$O$_3$ ($x$ = 0, 0.02, 0.05, 0.10, 0.15, 0.20) single crystals were grown by the optical floating zone technique using the recipe introduced in ref. [23]. Using the IP-XRD Laue Camera (TRY-IP-YGR, IPX Co., Ltd. Japan) and the high-resolution single-crystal X-ray diffractometer (XtaLAB P200, Rigaku Japan), we confirmed the high quality of the crystal[23] (see Fig. S3). Further characterization was done by measuring the field-cooled and zero-field-cooled DC magnetic susceptibility and AC susceptibility (MPMS-XL5 and MPMS-3, Quantum Design USA), which confirmed the long-range order without the spin-glass signature in $h$-YMn$_{1-x}$Al$_x$O$_3$. Some of the results are summarized in Fig. 1c. For inelastic neutron scattering experiments, the samples were cut into pieces with smaller sizes and were co-aligned on Al sample holders (2.2 g for $h$-YMnO$_3$, 3 g for $h$-YMn$_{0.95}$Al$_{0.05}$O$_3$ and $h$-YMn$_{0.9}$Al$_{0.1}$O$_3$, and 1.5 g for $h$-YMn$_{0.85}$Al$_{0.15}$O$_3$).

**Inelastic neutron scattering (INS) experiments.** We carried out INS experiments on the single crystal $h$-YMn$_{1-x}$Al$_x$O$_3$ using two time-of-flight (ToF) spectrometers: the MAPS spectrometer at ISIS, UK[39] for $h$-YMnO$_3$, and the 4SEASONS spectrometer at J-PARC, Japan[40] for $h$-YMn$_{0.9}$Al$_{0.1}$O$_3$ and $h$-YMn$_{0.85}$Al$_{0.15}$O$_3$. In the case of $h$-YMnO$_3$, the data were collected at 4 K with the incident neutron energy ($E_i$) of 30 meV. The chopper frequency was set to 350 Hz, which yields a resolution of 0.50 ~ 0.80 meV depending on the energy transfer, as shown in Fig. S4. For $h$-YMn$_{1-x}$Al$_x$O$_3$ with $x$ = 0.1 and 0.15, the data were collected at 5 K with multiple $E_i$ (6.8, 10, 16, 30, and 75 meV) and the Fermi chopper frequency of 250 Hz (see Fig. S4 for the instrumental resolution), thanks to the repetition-rate-multiplication (RRM) method implemented in 4SEASONS[41]. In all experiments, the samples were mounted in the geometry of (HHL) plane horizontal and were rotated during the measurement. For the data analysis, we used the Utsusemi[42] and Horace software[43]. Considering the crystal and magnetic symmetry of $h$-YMnO$_3$, the data were symmetrized into the irreducible Brillouin zone, which reduced the data's statistical error. Also, as the magnon modes' dispersion along the $c^*$-axis is negligible in $h$-YMn$_{1-x}$Al$_x$O$_3$, the data were integrated over the $c^*$-axis direction in a range of L= [−3, 3].

To acquire further information, we also carried out INS experiments in the Taipan triple-axis spectrometer at ANSTO, Australia, for $h$-YMnO$_3$ and $h$-YMn$_{0.95}$Al$_{0.05}$O$_3$. The data were collected at 5 K with the scattered neutron energy of 14.86 meV.

**Magnon linewidth analysis.** To estimate the $Q$-dependence of magnon energy linewidth, we performed magnon peak fittings at the $Q$ points within the full Brillouin zone. We used a Lorentzian function to fit magnon peaks, while a Gaussian function fitted the incoherent quasi-elastic signal. Instrumental resolution effects were removed from the fitted HWHM, assuming that the following relation holds:

$$\text{HWHM}_{\text{fit}} = \sqrt{(\text{HWHM}_{\text{instrument}})^2 + (\text{HWHM}_{\text{intrinsic}})^2}, \qquad (2)$$

where HWHM$_{\text{instrument}}$ was derived from the profile shown in Fig. S4. Note that for the doped samples, HWHM$_{\text{intrinsic}}$ is much larger than HWHM$_{\text{instrument}}$ at most $Q$ points (HWHM$_{\text{intrinsic}} \cong$ 10 HWHM$_{\text{instrument}}$), which guarantees the validity of Eq. 2. As there is more than one magnon peak at a certain $Q$ point, HWHM values of the magnon peaks were averaged at each $Q$ point, the results of which are displayed in Fig. 3d-3f.

**Theoretical calculations.** To take the dilution effect into account when performing spin-wave calculations, randomly distributed vacancies were created into a two-dimensional triangular lattice of 30 × 30 sizes with periodic boundary conditions. The resultant magnetic ground state affected by the vacancies was derived by simulated annealing followed by the conjugate gradient method. Using LSWT, we diagonalized the spin Hamiltonian (Eq. 1) with the ground state obtained from the previous process and calculated corresponding INS cross-sections using the SpinW library[44]. To get a statistically good result, we averaged the calculated INS cross-sections over 40 impurity replicas. We applied the same method to calculate the spin-wave spectra in a diluted square lattice antiferromagnet (see Fig. S9) and a diluted TLAF without spin texture. Although 120˚ magnetic order does not correspond to a classical energy minimum in a diluted TLAF, the validity of such examination still holds as far as the resulting magnon spectra are not ill-defined.

For precise comparison with the data, we performed energy and momentum resolution convolution based on the technical information of each ToF beamline (see Fig. S4). Notably, the effect of data integration over [00L] was included by calculating the average of the INS cross-sections with different L values (at 0.1 r.l.u. steps). The weighting factor of each piece was determined by the histogram of detector counts included in the data plot as a function of L. As a result, we confirmed a good agreement between the calculated magnon spectra and the data (Fig. S6).

## Acknowledgments

We thank Collin Broholm and Daniel I. Khomskii for fruitful discussions. This work was supported by the Leading Researcher Program of the National Research Foundation of Korea (Grant No. 2020R1A3B2079375) and the Institute for Basic Science in Korea (IBS-R009-G1). The inelastic neutron scattering experiments at the Japan Proton Accelerator Research Complex (J-PARC) was performed under the user program (Proposal and No. 2016A0171 No. 2017B0005). M.E.Z. acknowledges financial support from ANR, France (Grant No. ANR-18-CE05-0023). The work of A. L. C. was supported by the U.S. DOE, Office of Science, Basic Energy Sciences under Awards DE-FG02-04ER46174 and DE-SC0021221.## Author contributions

[†]These authors contributed equally to the work. J.-G.P. initiated and supervised the project. S.-W.C. and H.S. grew the single crystals and measured the bulk properties. K.P., J.O., J.C.L., T.K., J.J., K.C.R., K.K., K.I., T.G.P., and H.W. performed the inelastic neutron scattering experiments. K.P. and K.H.L. carried out the ground state calculations, and P.P., K.P., and J.O. did the spin-wave calculations. K.H.L., M.E.Z., and A.L.C. contributed to the theoretical interpretation and discussion. P.P., K.P., and J.-G.P. wrote the manuscript with contributions from all authors.

**Competing interests** The authors declare no competing financial interests.

**Supplementary Information** is available for this paper at [website url].

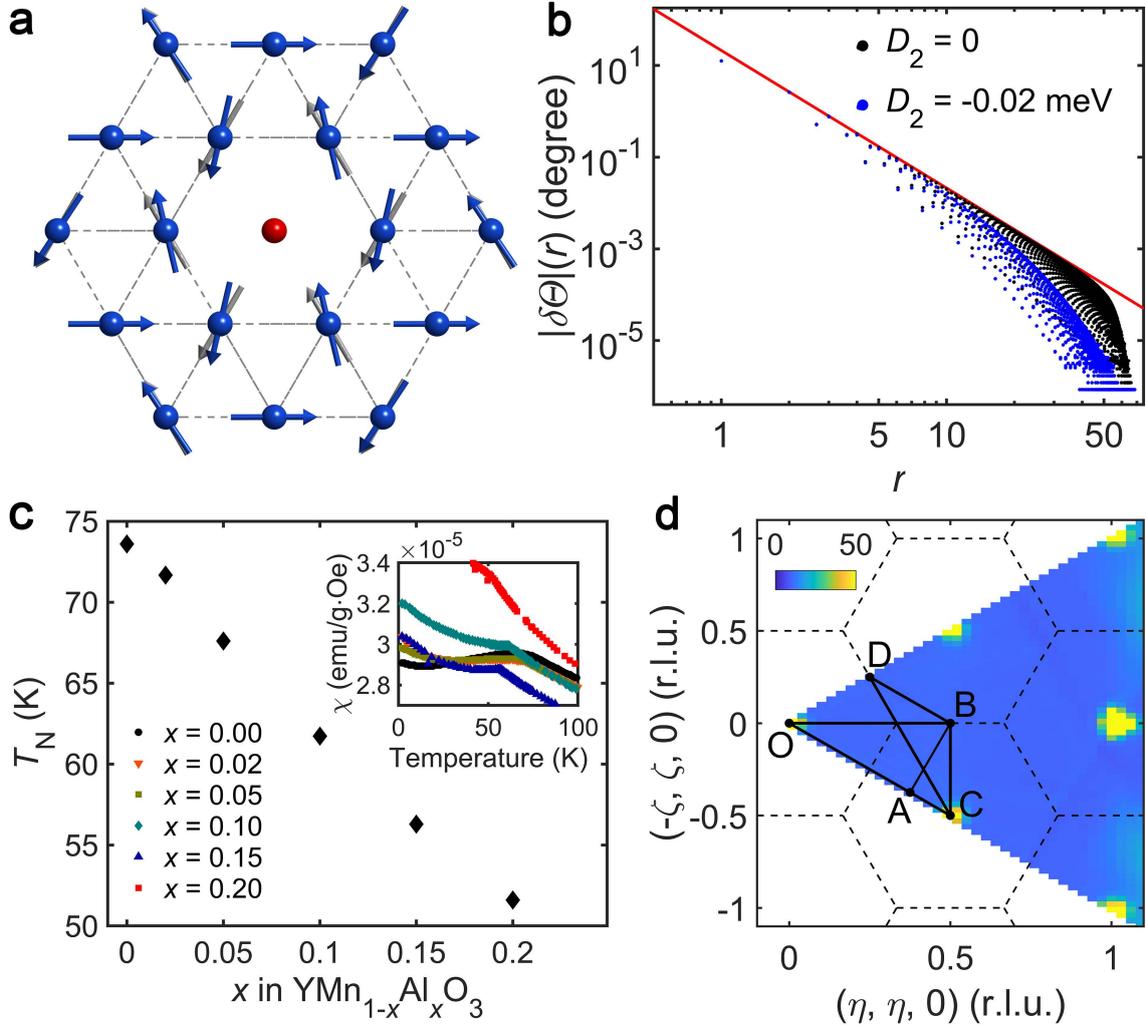

**Fig.1 | Formation of a spin texture in 2D-TLAF $h$-Y(Mn,Al)O$_3$. a**, Local spin canting induced by a vacancy in a triangular lattice antiferromagnet (blue arrows) and original 120° magnetic order (grey arrows). **b,** Distance dependence of the canting angle $|\delta\Theta(r)|$ in the spin texture around a vacancy determined from the numerical simulations. Black (Blue) dots denote the simulation with $D_2 = 0$ (0.02) meV, repetively. The red line is a guided plot of the $1/r^3$ behavior. **c,** Doping dependence of the transition temperature $T_N$. The inset denotes the magnetic susceptibility of YMn$_{1-x}$Al$_x$O$_3$ as a function of temperature, from which the transition temperatures ($T_N$) were determined. **d,** Two-dimensional reciprocal space of YMnO$_3$ with the labels of high-symmetric points and the high-symmetric lines used in Fig. 2.

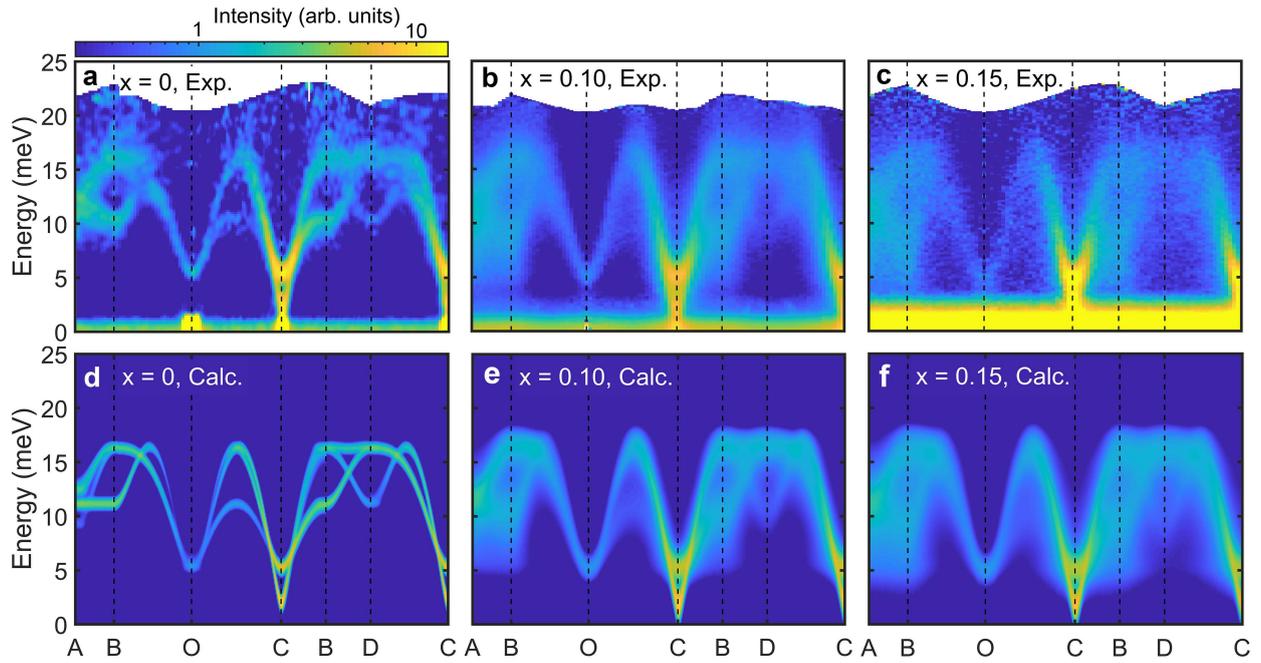

**Fig.2 | Magnetic excitation spectra of *h*-YMn$_{1-x}$Al$_x$O$_3$. a-c,** INS spectra of **a** *h*-YMnO$_3$ (ISIS), **b** *h*-YMn$_{0.9}$Al$_{0.1}$O$_3$ (J-PARC), and **c** *h*-YMn$_{0.85}$Al$_{0.15}$O$_3$ (J-PARC) measured at 5 K with the incident neutron energy of $E_i$ = 30 meV. The scattering intensity of the data was integrated over the $c^*$-axis. **d-f**, Theoretical INS cross-section of **d** *h*-YMnO$_3$, **e** *h*-YMn$_{0.9}$Al$_{0.1}$O$_3$, and **f** *h*-YMn$_{0.85}$Al$_{0.15}$O$_3$, which include the instrumental resolution convolution as well as the data integration effect over the $c^*$-axis (see Methods).

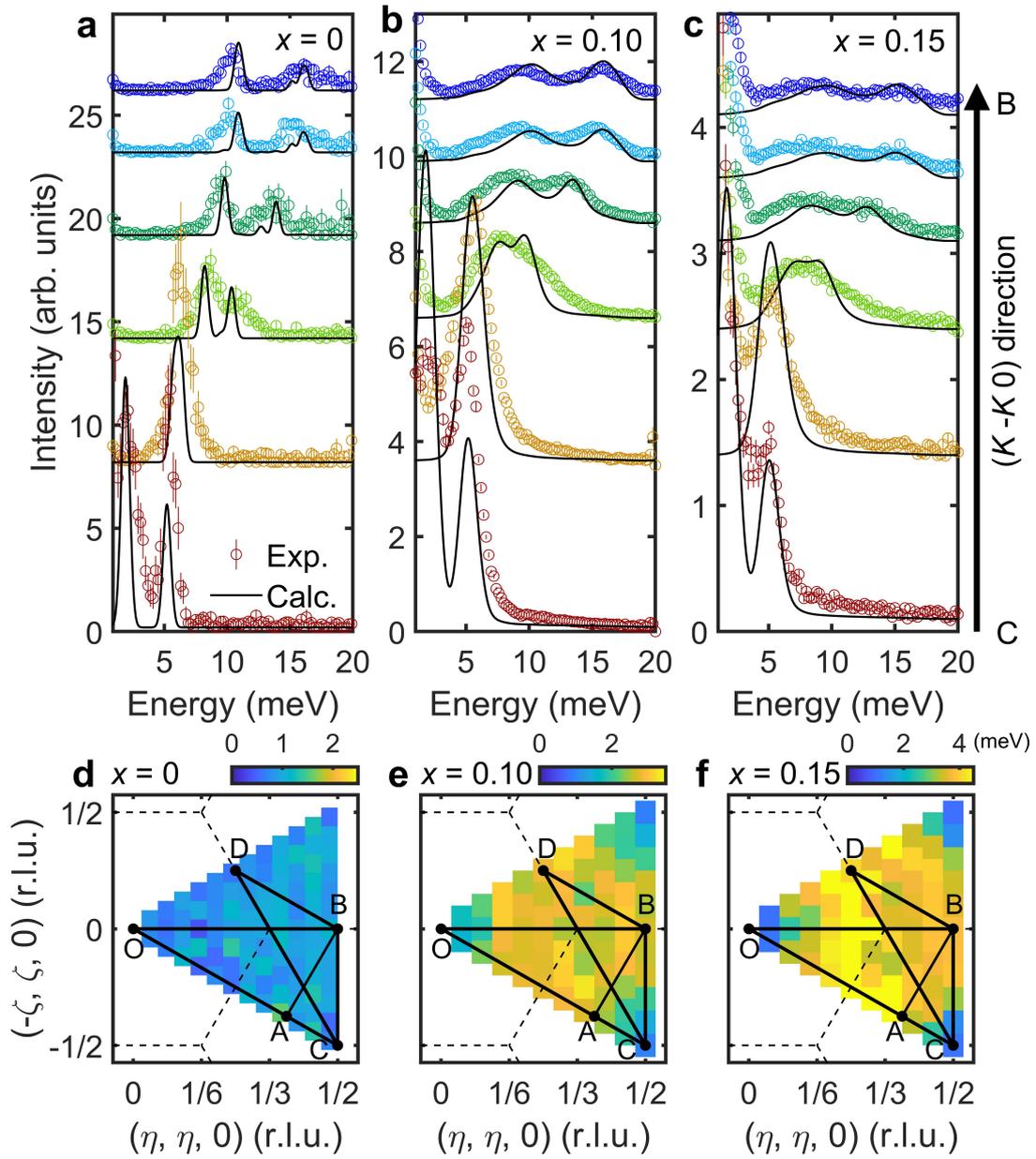

**Fig.3 | *Q*-dependent magnon linewidth broadening due to doping. a-c**, Constant *Q*-cuts (colored circles) and theoretically calculated INS cross-sections (solid black lines) of **a** *h*-YMnO$_3$, **b** *h*-YMn$_{0.9}$Al$_{0.1}$O$_3$, and **c** *h*-YMn$_{0.85}$Al$_{0.15}$O$_3$ at various *Q* points along the [K, -K, 0] direction (the C-B direction). Both the experimental data and the theoretical calculation results are rendered with the intensity integration along the *c*∗-axis. **d-f**, Fitted intrinsic HWHM (Γ(*Q*)) of the magnon modes in **d** *h*-YMnO$_3$, **e** *h*-YMn$_{0.9}$Al$_{0.1}$O$_3$, and **f** *h*-YMn$_{0.85}$Al$_{0.15}$O$_3$ over the full Brillouin zone, where the instrumental resolution effects are excluded (see Methods).

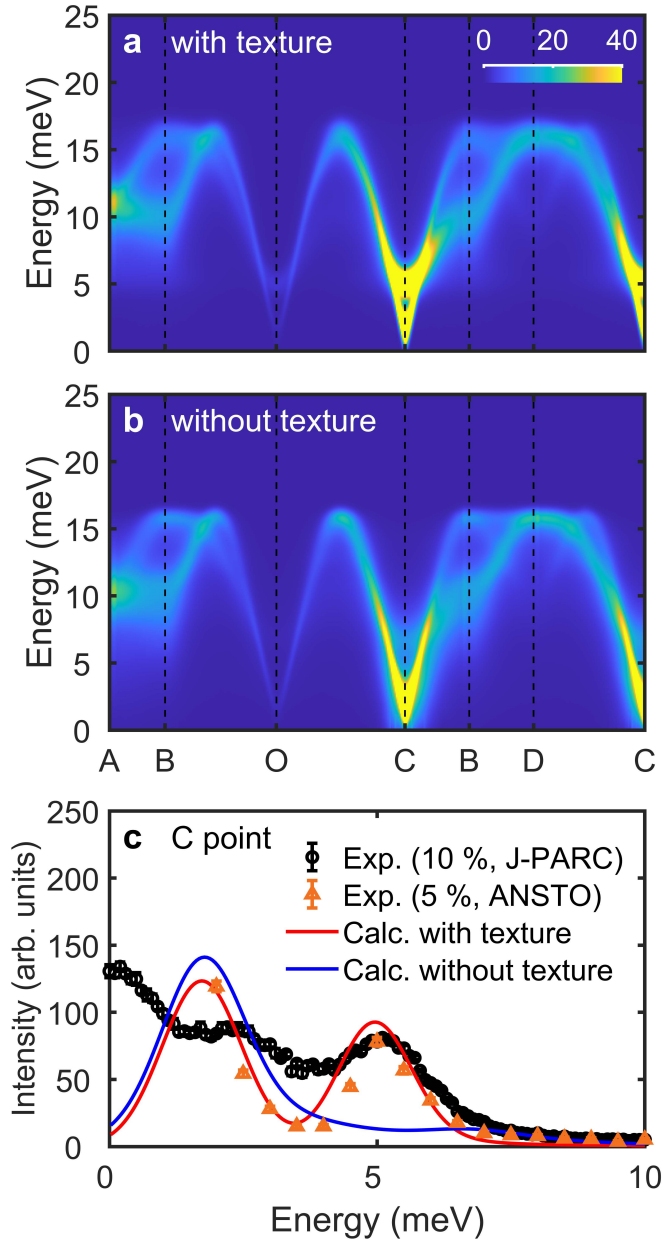

**Fig.4 | A role of spin textures in the spin dynamics of $h$-YMn$_{1-x}$Al$_x$O$_3$.** Theoretical spin-wave spectra of $h$-YMn$_{0.9}$Al$_{0.1}$O$_3$ **a** with vacancies and spin textures, and **b** with vacancies but without spin textures. **c**, A constant $Q$-cut at the C point, which demonstrates the suppression of the 5 meV mode in the calculation results without spin texture. Black and orange points are the INS data of $h$-YMn$_{0.9}$Al$_{0.1}$O$_3$ and $h$-YMn$_{0.95}$Al$_{0.05}$O$_3$, respectively. Blue and Red solid lines are the constant-$Q$ cuts of **a** and **b**. Note that this figure's calculation results do not include the effect of data integration along the $c^*$-axis.

# Supplementary Information: Spin texture induced by nonmagnetic doping and spin dynamics in 2D triangular lattice antiferromagnet *h*-Y(Mn,Al)O$_3$

Pyeongjae Park[1,2,3†], Kisoo Park[2,3†], Joosung Oh[2,3], Ki Hoon Lee[2,3,4,5], Jonathan C. Leiner[2,3,6], Hasung Sim[2,3], Taehun Kim[1,2,3], Jaehong Jeong[1,2,3], Kirrily C. Rule[7], Kazuya Kamazawa[8], Kazuki Iida[8], T. G. Perring[9], Hyungje Woo[9,10], S.-W. Cheong[11], M. E. Zhitomirsky[12], A. L. Chernyshev[13] & Je-Geun Park[1,2,3*]

[1]*Center for Quantum Materials, Seoul National University, Seoul 08826, Republic of Korea*
[2]*Center for Correlated Electron Systems, Institute for Basic Science, Seoul National University, Seoul 08826, Republic of Korea*
[3]*Department of Physics and Astronomy & Institute of Applied Physics, Seoul National University, Seoul 08826, Republic of Korea*
[4]*Center for Theoretical Physics of Complex Systems, Institute for Basic Science, Daejeon 34126, Korea*
[5]*Department of Physics, Incheon National University, Incheon 22012, Korea*
[6]*Physik-Department, Technische Universität München, D-85748 Garching, Germany*
[7]*Australian Nuclear Science and Technology Organisation, Lucas Heights, 2234 New South Wales, Australia*
[8]*Comprehensive Research Organization for Science and Society (CROSS), Tokai, Ibaraki 319-1106, Japan*
[9]*ISIS Pulsed Neutron and Muon Source, STFC Rutherford Appleton Laboratory, Didcot, Oxfordshire, OX11 0QX, United Kingdom*
[10]*Department of Physics, Brookhaven National Laboratory, Upton, New York 11973, USA*
[11]*Department of Physics and Astronomy, and Rutgers Center for Emergent Materials, Rutgers University, Piscataway, New Jersey 08854, USA*
[12]*Université Grenoble Alpes, CEA, IRIG, PHELIQS, 38000 Grenoble, France*
[13]*Department of Physics and Astronomy, University of California, Irvine, California 92697, USA*
* Corresponding author: jgpark10@snu.ac.kr

**Supplementary Note. Spin texture in (anti-)trimerized triangular lattice antiferromagnet (TLAF)**

As the spin dynamics of $h$-YMnO$_3$ can be well explained by the spin Hamiltonian without a trimerization effect[1], we did not consider trimerization in our model calculation. Haing said that, it is still worth investigating whether trimerzation in $h$-YMnO$_3$ gives noteworthy changes on the spin texture. Therefore, we examined the change of spin texture due to (anti-)trimerization by giving exchange interactions with different sizes to trimerized bonds and anti-trimerized bonds, respectively. Based on the previous studies[2,3] on the spin dynamics of $h$-YMnO$_3$, we chose $J_2 = 0.8J_1$ (a trimerized case) and $J_2 = 1.2J_1$ (an anti-trimerzed case).

Using the same procedure described in the Method of the main text, we calculated the canted ground state due to a non-magnetic impurity with (anti-)trimerization (anti-)trimerized TLAF. Fig. S2 shows the canting angle of spins $|\delta\Theta(r)|$ as a function of the distance from the impurity site ($r$) on a logarithmic scale, both with and without (anti-)trimerzation. The result demonstrates that the aspect of spin canting in (anti-)trimerized TLAF remains similar to that in ideal TLAF at the intermediate region with little finite size effects (r<30). Therefore, our model calculations without the trimeriation effect well represents the possible spin texture and correspondsing spin dynamics in $h$-YMn$_x$Al$_{1-x}$O$_3$.

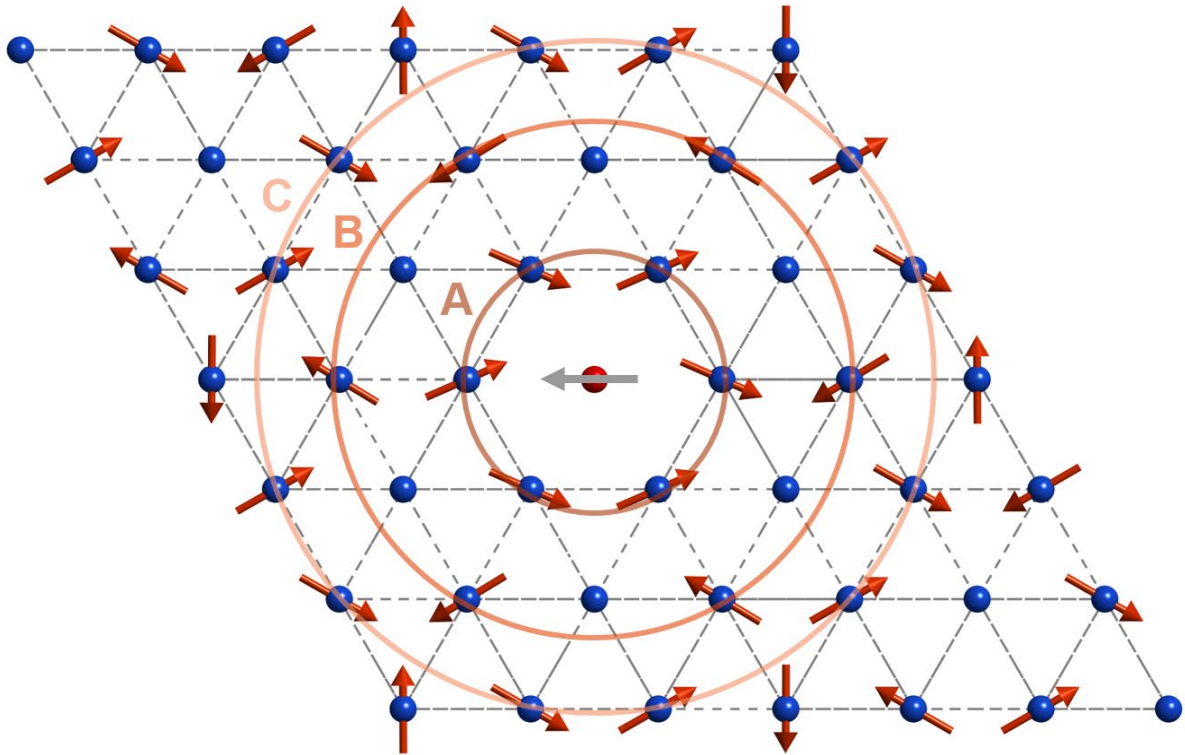

**Fig.S1** | The normalized difference between the spin configuration with and without spin texture, i.e., $\vec{S}_{canted} - \vec{S}_{120°}$. A gray arrow at the center denotes the net magnetic moment (-$S_0$) introduced by a non-magnetic impurity. Spins on the same circle (A~C) has the similar canting direction (left or right), which demonstrate the spatial structure of spin texture. Sites without a red arrow mean zero spin canting. Note that magnitude of the net difference at each site is normalized to the identical value for better demonstration.

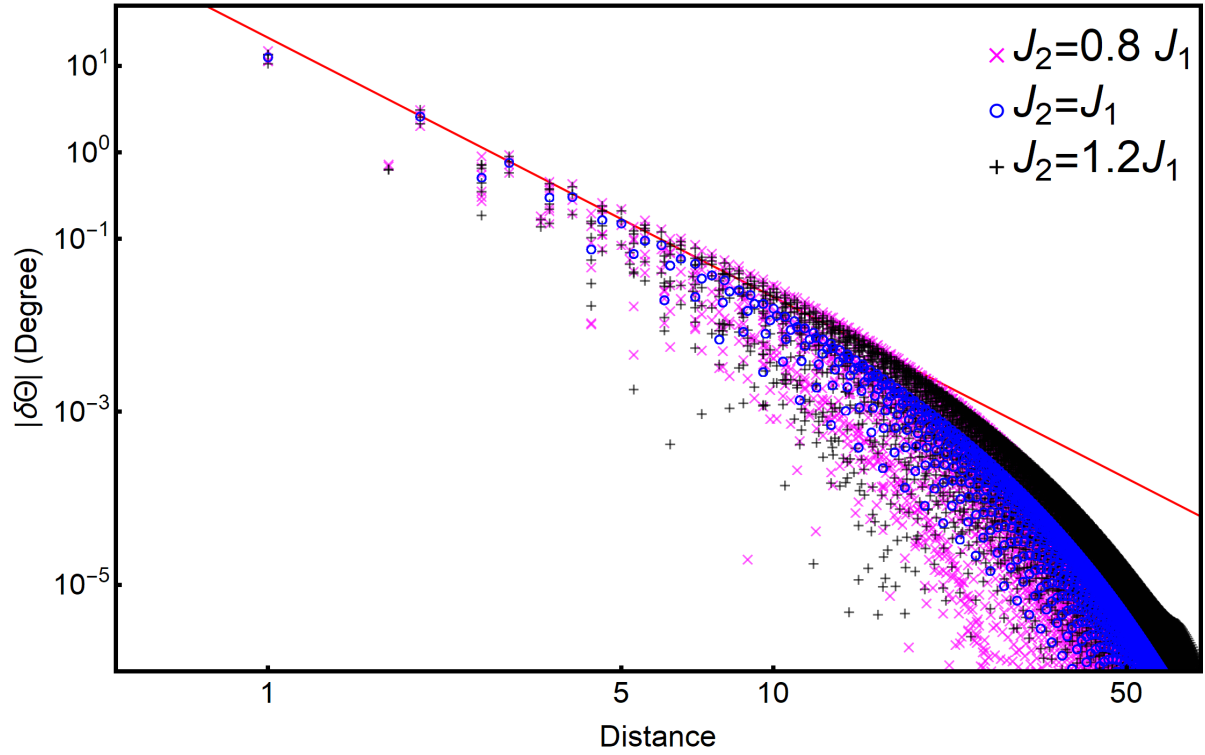

**Fig.S2** | Distance dependence of the canting angle $|\delta\Theta(r)|$ in the spin texture around a vacancy for an ideal (o), a trimerized (x), and an anti-trimerized triangular lattice antiferromagnet (+). The red line is a guided plot of the $1/r^3$ behavior. We used the notation of $J_1$ and $J_2$ same as those in refs. 1-3 of the supplementary information. The figure clearly shows that the spin texture of the (anti-)trimerized case remains similar to that of an ideal triangular lattice.

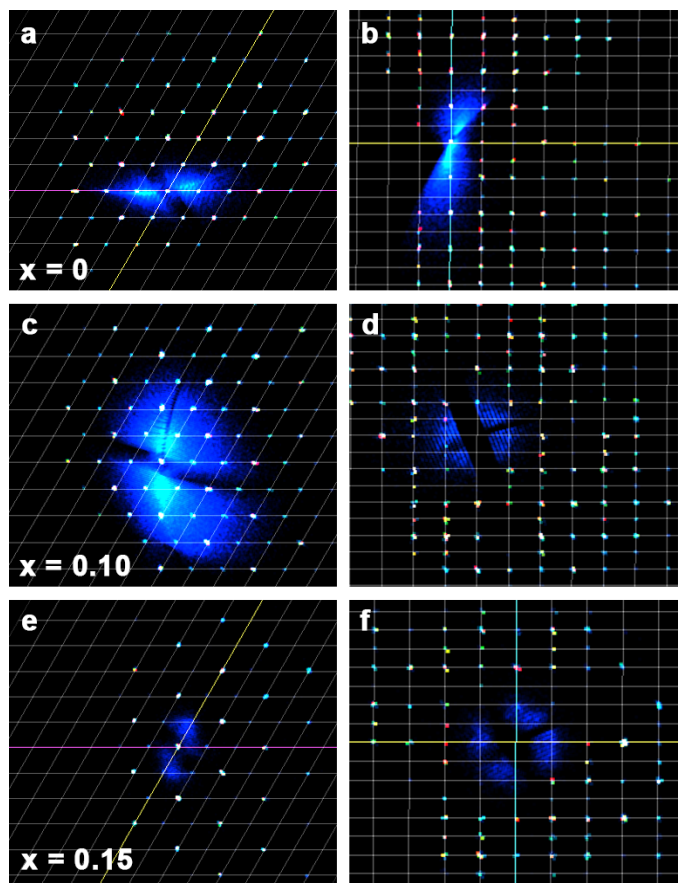

**Fig.S3 |** X-ray diffraction patterns of **a,b** h-YMnO3, **c,d** h-YMn$_{0.9}$Al$_{0.1}$O$_3$, and **e,f** h-YMn$_{0.85}$Al$_{0.15}$O$_3$ single crystals mapped on the a*-b* and b*-c* planes of the reciprocal space.

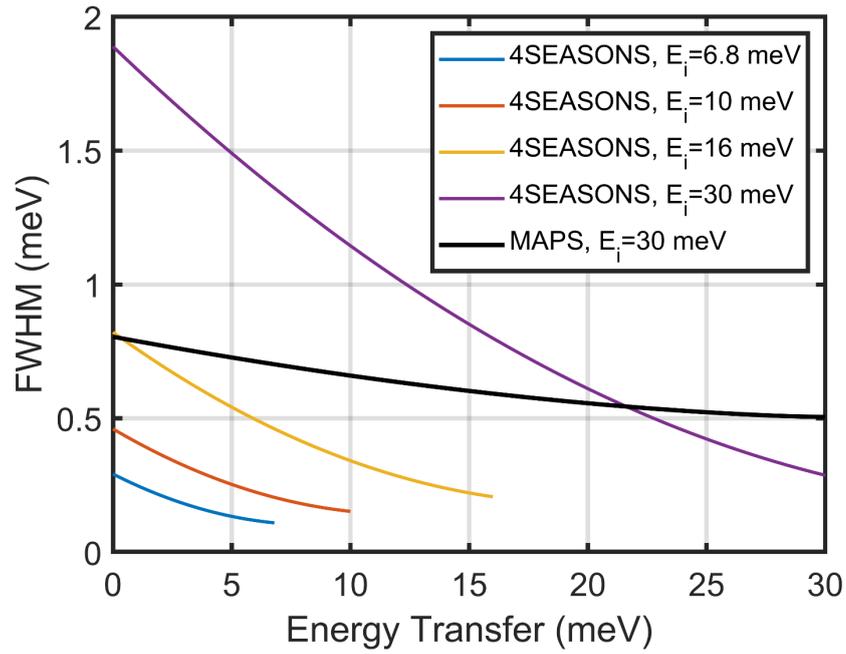

**Fig.S4** | Instrumental resolution of the time-of-flight neutron spectrometers used in this study. A solid black line denotes the instrumental resolution of the MAPS beamline with $E_i$ = 30 meV and a Fermi chopper frequency of 350 Hz. Colored lines are the instrumental resolution of the 4SEASONS beamline with $E_i$ = 6.8, 10, 16, 30, and 75 meV and a Fermi chopper frequency of 250 Hz when using the repetition-rate multiplication method.

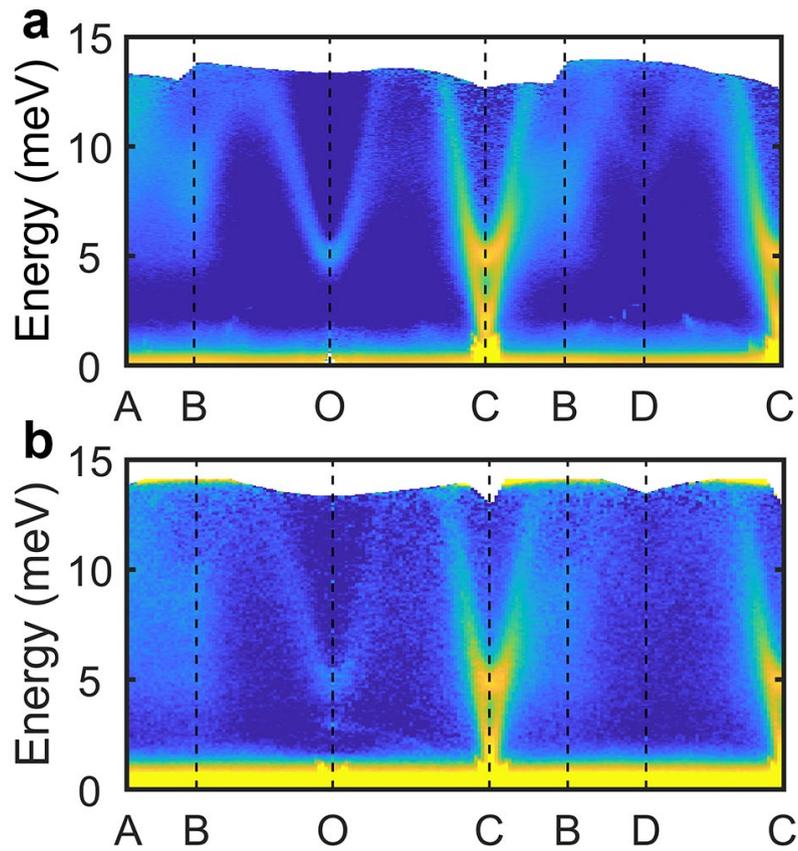

**Fig.S5** | INS spectra of **a** $h$-YMn$_{0.9}$Al$_{0.1}$O$_3$ (J-PARC), and **b** $h$-YMn$_{0.85}$Al$_{0.15}$O$_3$ (J-PARC) measured at 5 K with the incident neutron energy of $E_i$ = 16 meV. The scattering intensity of the data was integrated over the $c^*$-axis. Having better instrumental resolution than the data in Fig. 2, these data clearly show the severe broadening of the magnon modes at the zone boundary due to the doping, which is in stark contrast to the robustness of the magnon modes near the zone center (C point).

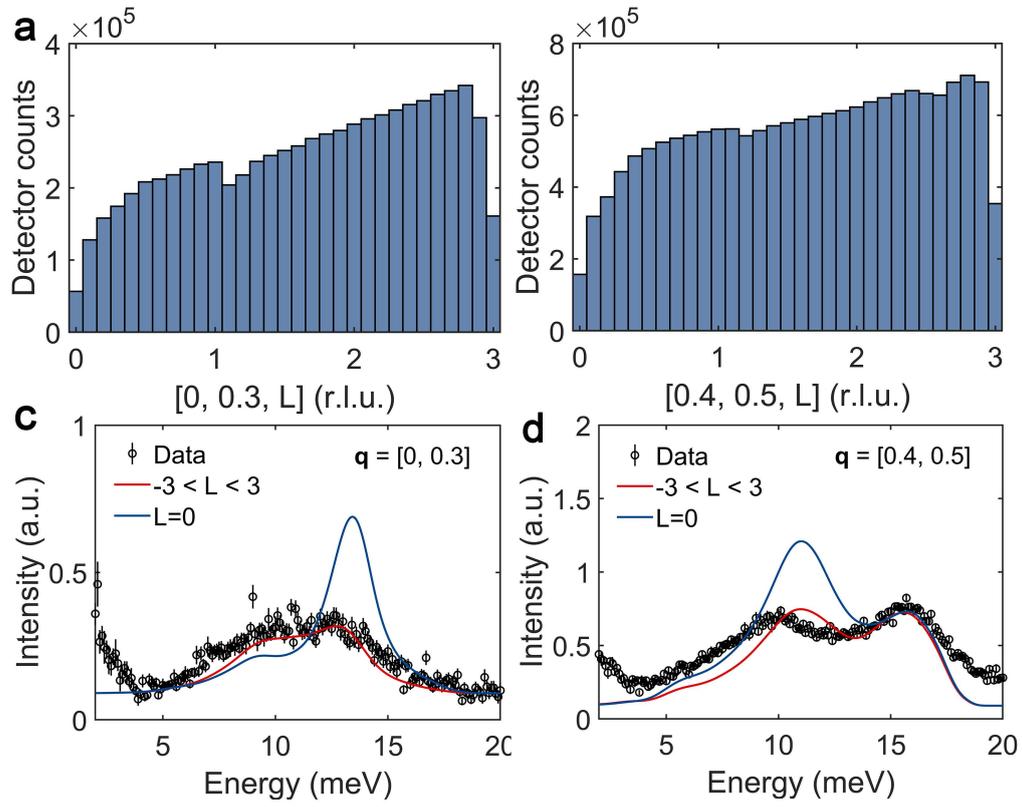

**Fig.S6 |** Data integration effects along the $c^*$-axis for $h$-YMn$_{0.9}$Al$_{0.1}$O$_3$. **a-b,** Histograms showing the individual momentum transfer of the detector counts included in the const-**Q** cuts at **a** [0, 0.3] and **b** [0.4, 0.5] (r.l.u.). They were used to consider the intensity integration along the $c^*$-axis in the calculation. **c-d,** Const-**Q** cuts at [0, 0.3] and [0.4, 0.5] (r.l.u.), and the calculated spin-wave spectra with (solid red lines) and without (solid blue lines) the data integration effect along the $c^*$-axis.

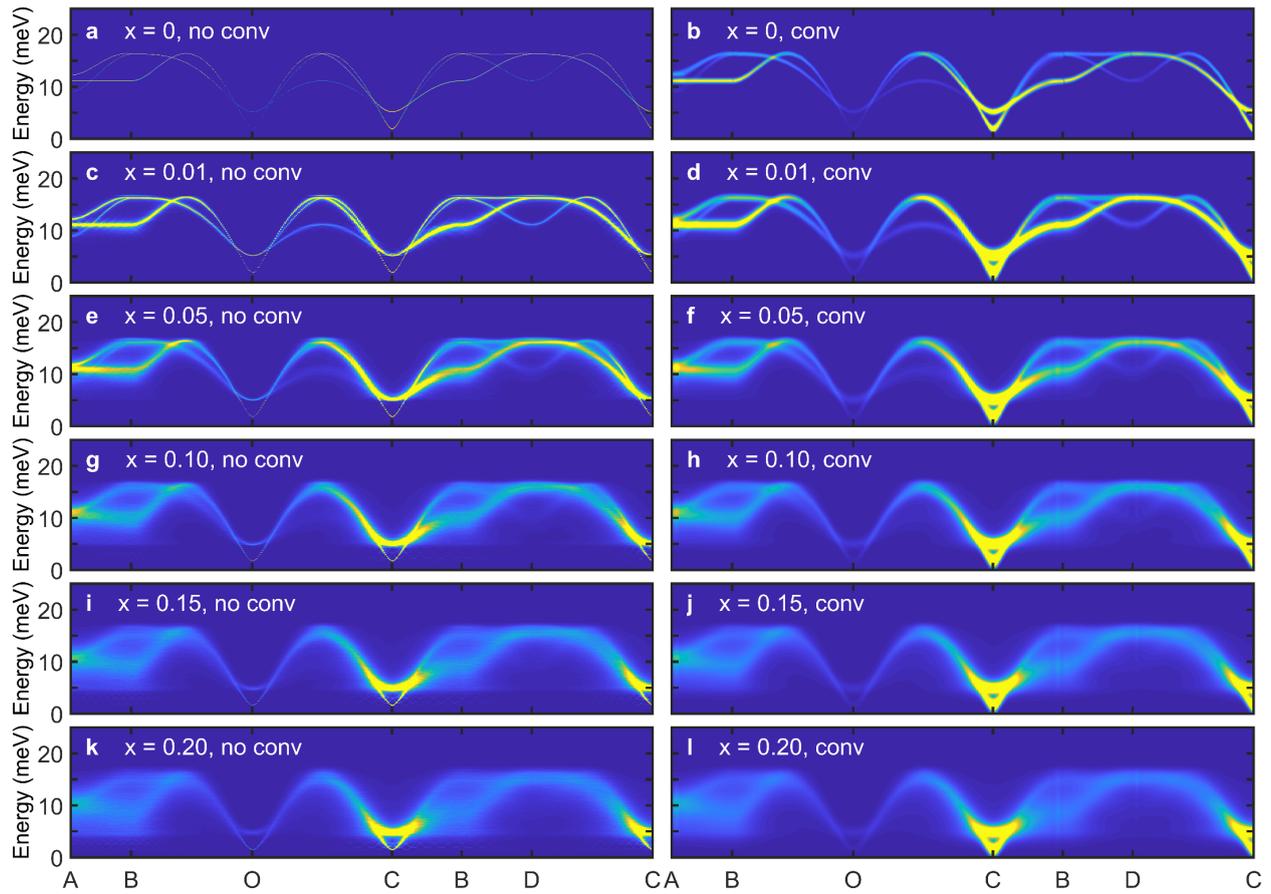

**Fig.S7 |** Theoretically calculated INS cross-sections of $h$-YMn$_{1-x}$Al$_x$O$_3$ with (a right panel) and without (a left panel) the instrumental resolution convolution. The color plots without the resolution convolution clearly show the intrinsic linewidth of magnon modes as well as the downward shift of the magnon dispersion along the A-B line. Note that the calculations in this figure do not include the data integration effect along the $c^*$-axis.

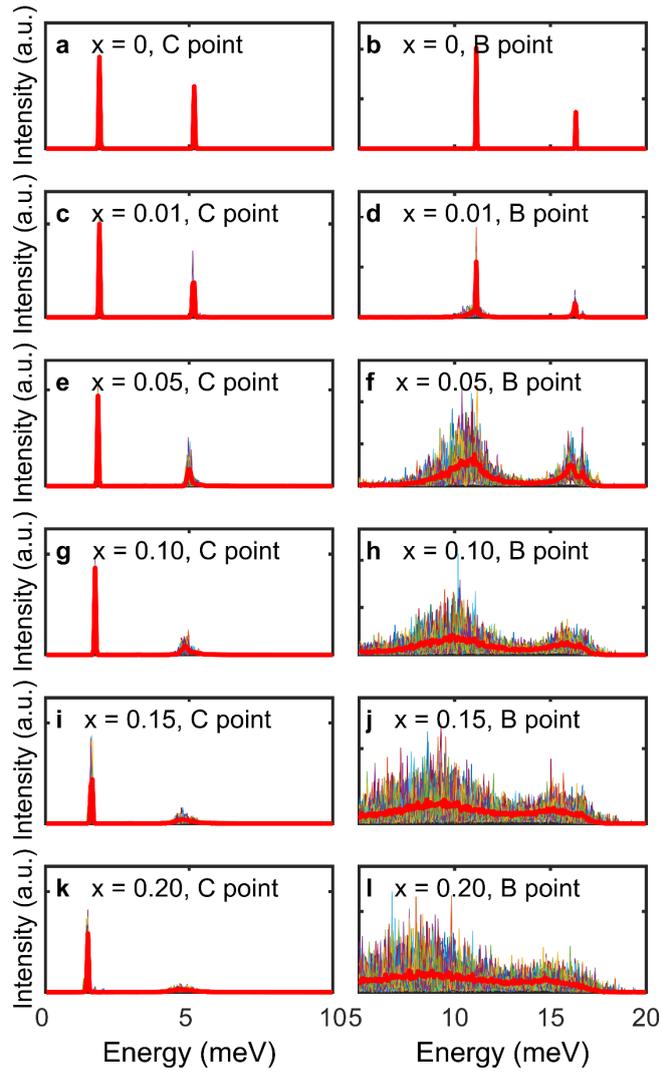

**Fig.S8 | a-c,** Constant-$Q$ cuts of the theoretically calculated INS cross-sections for $h$-YMn$_{1-x}$Al$_x$O$_3$ at the C point (left) and the B point (right). Thin lines are the calculated INS cross-sections from 40 replicas with random impurities (see Methods), which were used to calculate the ensemble average of the INS cross-sections. Red thick lines are the ensemble-averaged INS cross-sections of the 40 replicas. Comparing the results at the C point and the B point clearly show that the broadening effect under the doping is strongly $Q$-dependent. Note that the calculations in this figure do not include the data integration effect along the $c^*$-axis.

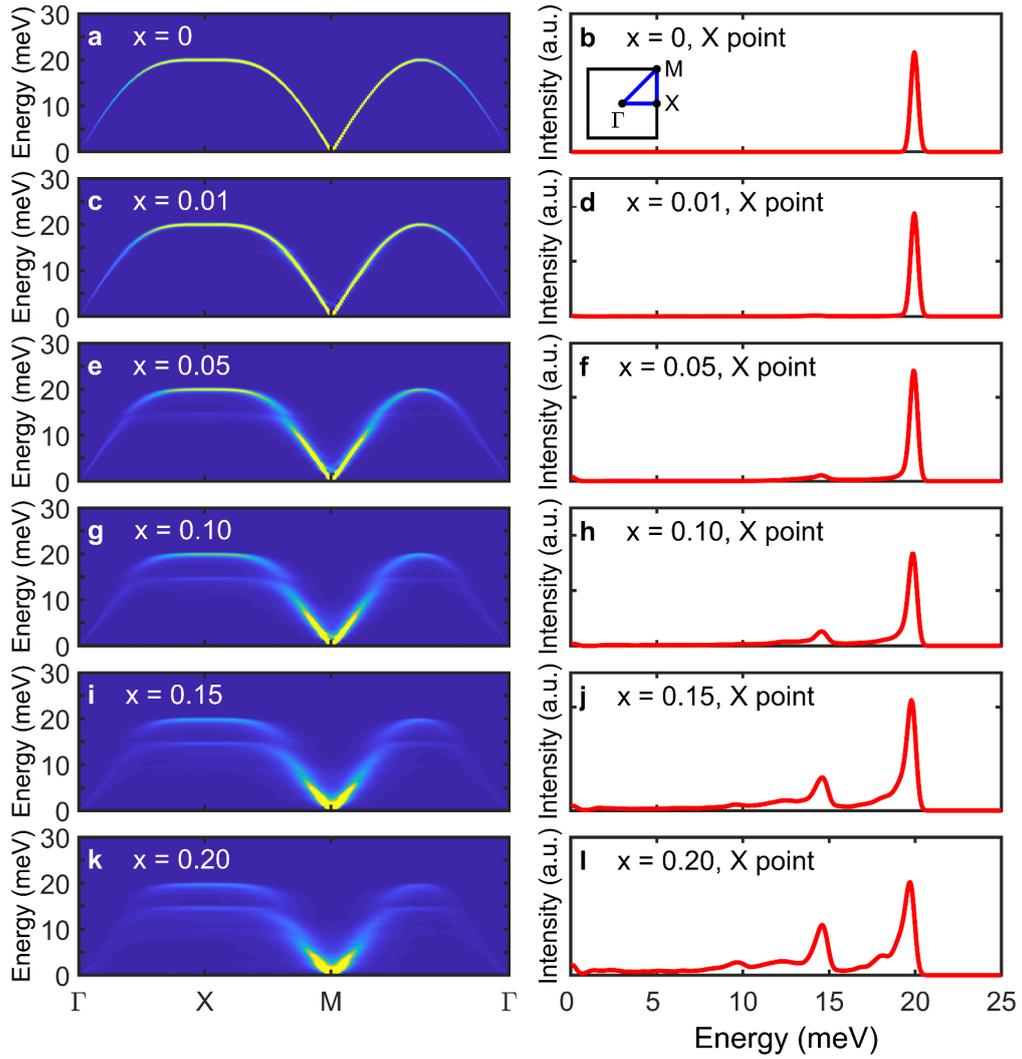

**Fig.S9 |** (Left) Calculated dynamical structure factor along the high symmetric lines in a diluted square lattice antiferromagnet with different amounts of doping. (Right) Constant *Q*-cuts of the results in the left panels at the X point.

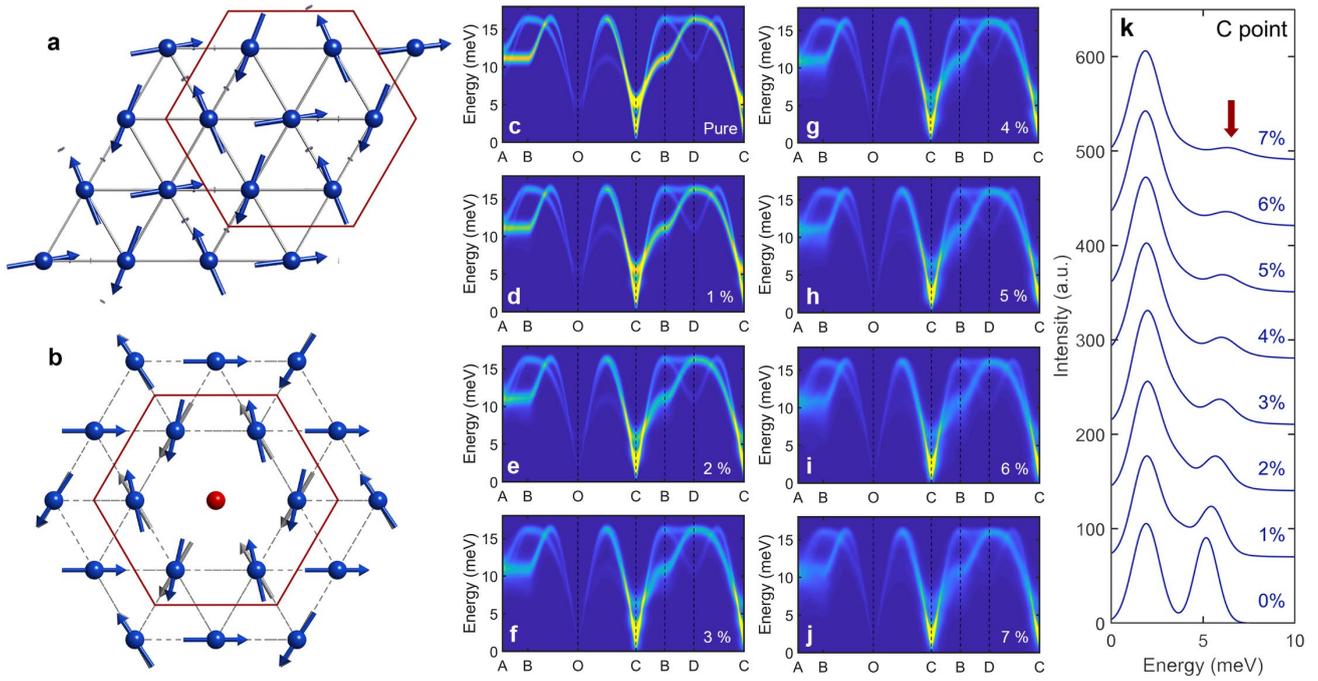

**Fig.S10 |** Selective vulnerability of the 5 meV magnon mode to the nonmagnetic impurity. **a,** A snapshot of the real-space spin precession for the 5 meV magnon mode at the C point. **b,** Spin canting due to the partial relief of the geometrical frustration by a nonmagnetic impurity (the same figure as Fig. 1a in the main text). Comparison between **a** and **b** shows the coincidence between the spin precession of the 5 meV mode and the spin canting due to the impurity. **c-j,** Doping dependence of the spin-wave spectra without considering the spin texture. **k,** Constant-$Q$ cuts of the calculation results in **c-j** at the C point, which demonstrates that the 5 meV mode (the red arrow) undergoes a significant change under the doping. Note that the calculations in this figure do not include the data integration effect along the $c^*$-axis.